Soft modes of the dielectric response in the twist-bend nematic ($N_{TB}$) phase and identification of the transition to nematic splay bend $N_{SB}$ phase in the dimer CBC7CB

K. Merkel[§1], A. Kocot[1], C. Welch[2] and G. H. Mehl[2]

[1]Institute of Material Science, Silesian University, Katowice, Poland
[2]Department of Chemistry, University of Hull, Hull HU6 7RX, UK

**ABSTRACT**

The dielectric spectra of the twist–bend nematic phase ($N_{TB}$) of (the bent-shaped) achiral liquid-crystal dimer 1''-,7''-bis(4-cyanobiphenyl-4'-yl) heptane (CB7CB) are studied for the determination of the different relaxation modes. Two molecular processes and one collective process were observed in the megahertz frequency range. The two molecular processes were assigned: one to the precessional rotation of the longitudinal components of the cyanobiphenyl groups and the second to the spinning rotation of the transverse component of the CB7CB dimer. The low frequency peak, at about 1MHz, shows a peculiar temperature behavior at the $N_{TB}$ to N transition, reminiscent of the soft mode at the transition from the SmA to the SmC phase. This peak can be assigned to a collective fluctuation of the tilt angle of the coarse grained director ***N*** with respect the pseudo-layer normal. This corresponds well with the electro-clinic effect observed as a response to an electric field in electro-optic experiments. The low frequency relaxation process, observed in the frequency range 1 Hz – $10^2$ Hz can be identified as a Goldstone mode, related to long-scale fluctuation of the cone phase. The birefringence data in the presence of strong bias fields in the temperature range where the $N_{TB}$ phase is formed is interpreted as unwinding of a helix and an indication of the formation of a field induced nematic splay bend phase ($N_{SB}$).

**Introduction**

In recent years, the has been a renewed interest in dimeric liquid crystal materials and bimesogens, mainly due to their extraordinary flexoelectric [1-3] and electro-optical properties [4-7] and for some dimeric materials the ability to form a modulated nematic phase ($N_{TB}$) [8-14] has led to increased attention. In the $N_{TB}$ phase, the director exhibits periodic twist and bend deformations forming a conical helix with doubly degenerate domains of opposite handedness [3,15,16]. The $N_{TB}$ phase in dimers was experimentally observed, mainly by a combination of experimental techniques such as polarized light microscopy, X-ray diffraction and Resonant Soft X-Ray Scattering (XRD, RSoXS) [5,9,11,17-21], light scattering [10] and freeze-fracture transmission electron microscopy (FFTEM) [5,9,11]. Additionally, the polarity and chirality of the $N_{TB}$ phase formed by the achiral molecules has been investigated by electro-optical [4-7,12,22], magneto-optical [23,25], dielectric [22,25-28] and nuclear magnetic resonance studies (NMR) [29-31]. A number of theoretical models have attempted to describe the new nematic ground state and the anomalously low or negative bent elastic constants [32-34]. All these methods suggest a local helical structure of extremely tight pitch, in the region of 8 -15 nm for dimers. Such nanoscale modulation of molecular orientation promises extremely short response times. Indeed the time of electro-optic switching reported for bent-core is about 1μs [25] and for CB7CB and CB11CB dimers even shorter times being in the sub-microsecond range have been detected, much shorter [3,4,7] when compared to the ms range for uniaxial nematics. The field-induced distortional effect in the $N_{TB}$ phase may be extremely useful for technological applications as this has similarity to the electro-clinic effect [4,35]. In the $N_{TB}$ phase, the effect of an electric field can be observed through strong dielectric coupling as well as by a weak flexoelectric effect [36,38]. A linear electro-optic effect is observed where the optical axis is rotated by the field applied in a plane perpendicular to the helical axis. Such a characteristic effect resembles the electroclinic effect, (ECE), in the SmA* phase [3] and flexoelectric effect (FEE) [35] in the N* phase. The analogy of $N_{TB}$ phase to N* phase is quite direct, as effects in both phases arise from flexoelectricity and both of these occur by distortions in the helical structure produced by the electric field. The analogy with the ECE in the chiral SmA phase (SmA*) is much more subtle. It can be understood in terms of the generalized analogy between the pseudo-lamellar $N_{TB}$ phase and lamellar SmA* [38].

The main motivation of this paper is to analyze the molecular orientational dynamics of the $N_{TB}$ mesophase formed by the polar symmetric dimer CB7CB and to study the behavior of this phase under an electric-field. Dielectric spectroscopy under DC bias is used to investigate its response to an electric field. We demonstrate that due to the periodic structure of the $N_{TB}$ phase, the electro-optic effects are not nematic-like but are very similar to those in the smectic and cholesteric phases [39]. Reviewing the latest dielectric research for the CB7CB and related systems [22,26-28], we note that until now no one has reported the existence of a soft mode in LC dimers.

**Experimental**

The chemical structure of the 1"-,7"-bis(4-cyanobiphenyl-4'-yl) heptane (CB7CB) is given in Figure 1, the synthesis has been reported recently [4,40]. Wide band dielectric spectroscopic experiments in the frequency range from 0.1 Hz to 100 MHz were carried out on a planar-aligned cell with and without the bias field. The cell-spacing was varied from 1.6 µm to 10 µm. For the higher frequency dielectric measurements, gold plated cells filled with the material in the isotropic state were used. The alignment layers on the electrodes of the cell were created by coating and annealing the surfactants on the electrodes, prior to the cell's assembly. The amplitude of the probe field was in the range from 0.01 to 1V/µm, whereas the DC bias field up to 10 V/µm was applied for studying the electric field dependence of the soft mode. The real ($\varepsilon'$) and imaginary ($\varepsilon''$) parts of the complex permittivity were measured for both planar and homeotropic aligned cells under slow cooling from the isotropic state. In order to determine the dielectric amplitude, $\delta\varepsilon_j$, and the relaxation times, $\tau_j$, of each relaxation mode, the dielectric spectra were analyzed using the Cole-Cole Eqn. (1) that expresses the complex permittivity in terms of the various relaxation processes:

$$\varepsilon^*(\omega) - \varepsilon_\infty = \sum_{j=1}^{n} \frac{\delta\varepsilon_j}{[1 + (i\omega\tau_j)^{1-\alpha}]}, \qquad (1)$$

Here $\delta\varepsilon_j$, $\tau_j$, $\alpha_j$, $\varepsilon_\infty$ are the fitting parameters of the equation to the experimental data of $\varepsilon''$. Results show that the high frequency dielectric spectra (frequency range above 0.1 MHz) are much more complicated than expected for the rigid core molecules. CBnCB molecules are highly flexible and we expected contributions to the spectra not only from rotations of the components of the total molecular dipole moment but also coming from rigid cores of the mesogens.

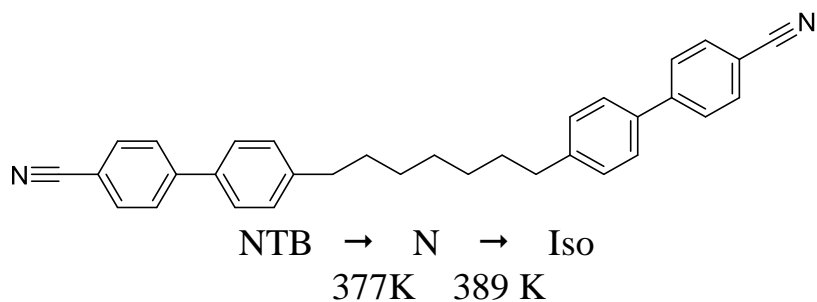

NTB → N → Iso
377K    389 K

**Fig.1.** Chemical structure of the dimer CBC7CB and transition temperatures [6].

The contributions of conformational changes to the $\varepsilon''$ spectra from parts of the mesogenic groups can also be detected. For obtaining a better deconvolution of the relaxation peaks in the frequency range above $10^5$Hz, it is preferable to analyze the derivative of the real part of permittivity $\varepsilon'$ [41] as given below:

$$\frac{d\varepsilon'}{d(\ln f)} = \frac{d\varepsilon'}{d(\ln \omega)} = \sum_{j=1}^{n} \text{Re} \frac{\delta\varepsilon_j \alpha (i\omega\tau_j)^\alpha}{[1+(i\omega\tau_j)^\alpha]^2}, \qquad (2)$$

**Results and discussion**

Analysing the temperature behaviour range below 376 K we can consider up to three relaxation peaks ($n \leq 3$), in the frequency range above 0.1 MHz and one in the low frequency range ~10 Hz. In the frequency range above 0.1 MHz the various modes: **m1**, **m2** and **m3** were obtained by fitting the experimental data to Eq. (1). The lowest frequency peak, **m4**, was analyzed separately as it is not overlapping with the others. A set of relaxation frequencies so obtained, **m1**, **m2** and **m3**, are plotted vs. temperature in Fig. 2a. The corresponding relaxation frequencies, $f_j = 1/2\pi\tau_j$, are calculated from $\tau_j$. From these the values for $\alpha$ are found to be in the range 0.05 - 0.1. The corresponding amplitudes of modes **m1**, **m2** and **m3** are plotted in Figure 2b. For the temperature range above 376 K, the spectrum is even more complex. The additional peak **m2'** which is due to the presence of other conformers can be deconvoluted. This peak can be assigned to the contribution of a hairpin-shaped conformer, likely to be present in the spectrum.

**Molecular modes**

The complex permittivity expressed in terms of the Maier and Meier (M-M) model as given by Toriyama et al [42] is used here. The dynamics of the dielectric relaxation in the nematic phase is usually interpreted in terms of the rotational diffusion model for the reorientation of molecules in the nematic field, the model given by Coffey and Kalmykov [43]. The perpendicular component of the complex permittivity, $\varepsilon_\perp^*(\omega)$, can be written as:

$$\varepsilon_\perp^*(\omega) - \varepsilon_{\perp\infty} \approx \frac{N'hF^2}{3\varepsilon_0 k_B T}\left[\frac{\mu_l^2(1-S)}{1+i\omega\tau_{10}} + \frac{\mu_t^2(1+\tfrac{1}{2}S)}{1+i\omega\tau_{11}}\right] \qquad (3)$$

$A = N'hF^2/3\varepsilon_0 k_B$ is the scaling factor for the two relaxation mechanisms that contribute to the complex permittivity. In Eq. (3), $N'(= M/d)$ is the number density of molecules, $d$ is the mass density and $M$ is the molecular weight, $\varepsilon_0$ is the permittivity of vacuum, $T$ is the absolute temperature, $k_B$ is the Boltzmann constant. $F$ and $h$ are the internal field factors for the reaction and cavity fields, respectively. $\mu_l$ and $\mu_t$ are the longitudinal and the transverse projections of the molecular dipole moment, $\mu$, directed along and normal to the long molecular axis, respectively. Thus, depending on the nature of the relaxation mode, the two terms on the right-hand side of Eq. (3) relax at different frequencies, $f_j=1/2\pi\tau_j$. In the rotational diffusion model [43], $\tau_{10}$ and $\tau_{11}$ are the relaxation times for the precessional and spinning rotations, respectively and are so assigned. Expressions for $\tau_{10}$ and $\tau_{11}$ are related with the relaxation time in the isotropic state, $\tau_0$. The orientational order parameter, $S$, and the anisotropy in the rotational diffusion coefficients $\Delta = \frac{1}{2}\left[D_\parallel / D_\perp - 1\right]$ are expressed by Eqs. 4a and 4b. $D_\parallel$ and $D_\perp$ are parallel and normal components of the rotational diffusion coefficients, respectively.

$$\frac{\tau_{10}}{\tau_0} = \frac{1-S}{1+\frac{1}{2}S} \qquad \frac{\tau_{11}}{\tau_0} = \frac{2+S}{2+\Delta(2+S)-\frac{1}{2}S} \qquad \text{(4a and 4b)}$$

Two relaxation peaks are dominant in the higher frequency range of the $\varepsilon''$ spectra, and these are assigned to the molecular relaxation modes **m1** and **m2** of the CBC7CB dimer dipole moments as reported by M. Cestari *et al* [18] and D. O. López *et al* [27]. Longitudinal, $\mu_l$, and transverse, $\mu_t$, components of the molecular dipole moment, $\mu$, contribute to the dielectric permittivity differently and they relax at different frequencies of the probe field. The corresponding relaxation peaks are found at the frequencies $f_{m2}$=8 MHz and $f_{m1}$=70 MHz (at 377 K) for transverse and longitudinal components respectively. The dielectric strengths of those two peaks, Fig.2(a) and their relaxation frequencies, Fig.2(b) are in good accordance with the earlier studies [18,27] in the range of $N_{tb}$ phase. Additionally, the other peak, corresponding to the **m3** mode, contributes the relaxation spectra of $\varepsilon''$, at the frequencies of about $f_{m3}$=1MHz, which has not been reported before.

In the $N_{TB}$ phase the contribution of the bent conformation is dominant. As the dimer is symmetric, the longitudinal component of its dipole is zero and the contribution to $\varepsilon_\perp$ is only due to the spinning rotation of the transverse component of the dimer dipole moment. Thus the second relaxation, which must be due to precessional rotation of the longitudinal component, can arise only from a rotation of a part of the dimer, i.e. the cyanobiphenyl mesogenic core alone. In the uniaxial N phase the frequency of the spinning motion cannot be unambiguously established because of significant contributions of other conformers. This is reflected by an increase of the strength of transverse component, $\delta\varepsilon_t$, at the transition from the $N$ to $N_{TB}$ phase, where the contribution from the bent conformation becomes dominant. Thus in the uniaxial N phase, the results are quite complicated for analysis as there are more conformers that contribute to the dielectric spectra because dynamics are strongly affected by conformational reorganizations of the molecules. In particular, an additional peak **m2'** can be deconvoluted in the uniaxial N phase due to the presence of other conformers, i.e. the hairpin-shaped conformer. On the transition to the *N* phase a significant decrease of $\delta\varepsilon$ can be clearly seen, which is due to a reduction of the population of bent conformers (for other conformers the transverse component can be neglected). For the same reason the contribution of the **m3** mode becomes smaller, as it is related to longitudinal component of the dimer dipole moment. Another important reason for discrepancy between our results and those of M. Cestari *et al* [18] and D. O. López *et al* [27] is the method of sample alignment. The relaxation time, for **m2** mode is measured for homeotropic alignment but sample alignment has been achieved in this case by applying strong electric field. This, however, is very likely to result in an increasing population of hairpin-shaped conformation in *N* phase, which are expected to have a different dynamic behavior than the bent core conformation (in our case relaxation times of **m2'** mode coincide with those reported by M. Cestari *et al* [18] and D. O. López *et al* [27]. In the low-temperature nematic phase, the dielectric anisotropy [$\triangle\varepsilon = \varepsilon_{\parallel} - \varepsilon_{\perp}$] becomes close to zero, and may change sign from positive to negative at transition temperature.

The dynamics for **m2** mode, i.e. its relaxation times are dependent on the order parameter, *S*, anisotropy of diffusion coefficient, $\triangle$, (3b). We can well reproduce the relaxation time $\tau_{11}$ by using, *S* parameter, obtained from diamagnetic anisotropy [44], or birefringence measurements [44,45]. We note that in [44,45] a classical Haller formula for the approximation the order parameter, *S,* was used. As result of our approach the anisotropy of diffusion coefficient, $\triangle$, can be obtained, Fig. 2b. The temperature dependence of the diffusion coefficient anisotropy, $\triangle$, see Fig.2b, clearly shows acceleration of the spinning rotation on transition from the *N* to the $N_{TB}$ phase, presumably due to enhancing the rotational freedom in the $N_{TB}$ phase. The relaxation time of the **m2** mode, on the other hand, can be related to rotational viscosity, $\gamma$, using formula:

$$\tau = 3\gamma V\lambda / k_B T, \tag{5}$$

where *V* is the volume per molecule of a LC dimer and $\lambda$, is the Perrin friction factor of asymmetric ellipsoid for its spinning rotation about the longest axis [47]. The factor, $\lambda$, was calculated to be 0.32 for the dimer molecule approximated by an ellipsoid with an axes ratio of 1:0.35:0.25, Fig.2. Thus both dynamic parameters the viscosity coefficient and the diffusion coefficient indicate an acceleration of the rotation on the transition to the $N_{TB}$ phase.

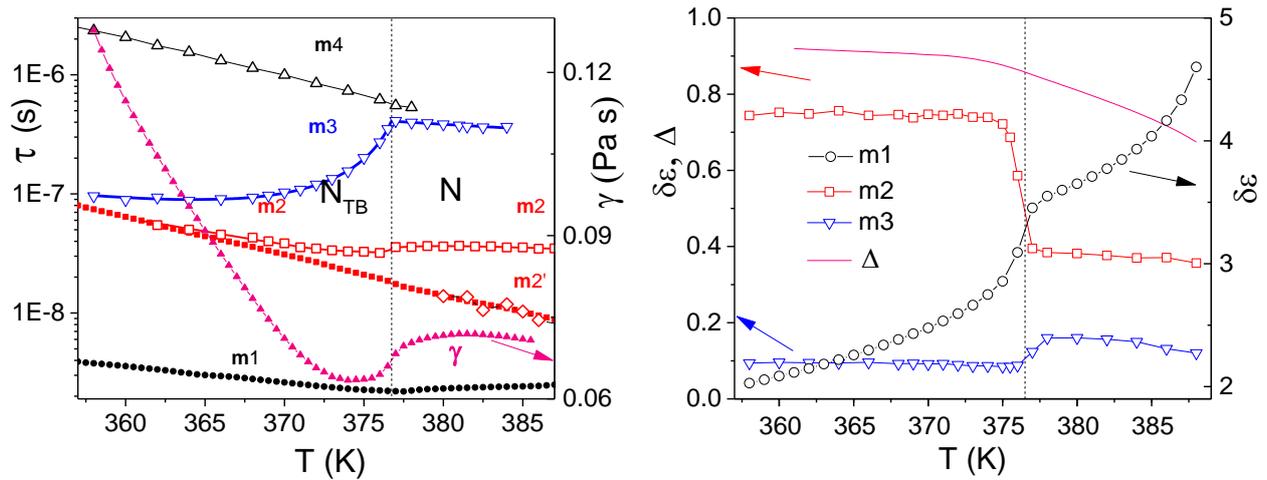

Fig. 2. (a) Plots of relaxation times for modes: **m1**-●  and **m2**-■, reported by López et al [27] and present results for modes: **m2**-□,  **m2'**-◇,  **m3** - ▽,  **m4** - △ (**m4** is multiplied by 3·10⁻⁵), plotted as a function of temperature, a continuous blue line is a fit to experimental data. The viscosity coefficient, $\gamma$ - ▲, calculated from relaxation time of the **m2** mode. (b) Plots of the the dielectric amplitude $\delta\varepsilon$, of the modes **m2**- ○,  **m2** - □, and **m3** - ▽. The anisotropy in the rotational diffusion coefficients, $\Delta$, is plotted as 'purple continuous line'.

**Collective modes**

The observed peak in the MHz frequency range, named **m3** mode, shows quite a specific temperature behaviour with a peculiarity at the $N_{TB}$ to N transition. This is reminiscent of the **soft mode** at the transition from the SmA to the SmC phase, see Fig 3. This peak can be assigned to a collective fluctuation of the tilt angle of the coarse grained director $N$ out of the pseudo-layers normal, similar what has been reported for bent-core molecules [25]. As a response to an electric field one can expect electro-optic effects similar to the in-plane flexoelectric switching in the N* phase. Such an electro-optic effect has been already observed for CB7CB dimer by Panov et al. [4] and C. Meyer et al [45]. A more extensive theoretical and experimental study by C. Meyer et al. [7] showed that this effect is also similar to the electroclinic effect (ECE) in the SmA* phase.

The $N_{TB}$ and the cholesteric phases have some important common features: both phases are periodic and have a pseudo-layered structure, and both of them are chiral (the cholesteric due to the molecular chirality and the twist-bend nematic because of the structural chirality of the heliconical structure). Therefore, in all three phases (N*, SmA* and $N_{TB}$), when an electric field **E** is applied in the plane of the layers (or pseudo-layers), the optic axis rotates from its initial orientation, along the normal to the layers, to a new orientation, which lies in the plane of the layers. This effect is only possible in chiral systems, thus confirming the chiral symmetry of the $N_{TB}$ phase even though it is formed by achiral molecules.

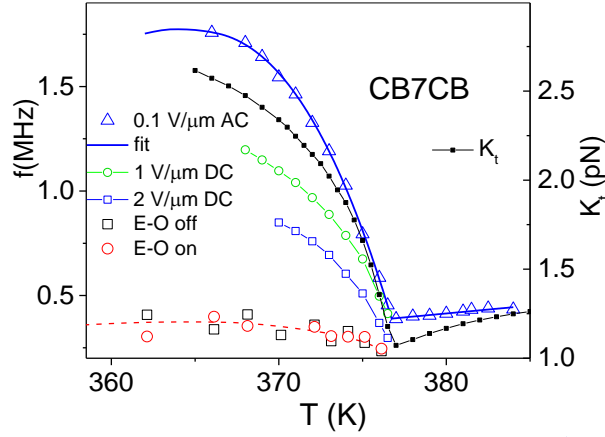

Fig. 3. Relaxation frequency of the tilt deformation mode, **m3**: $f_t$ - △. A continuous blue line is a fit to the experimental data and $K_t$ is elastic modulus obtained from the fitting, using eq.(5) -■- . Relaxation frequencies of the helix tilt deformation mode under DC field E: 1V/μm- ○, and 2 V/μm. The temperature dependence of the relaxation frequency corresponding the electroclinic ffect in the $N_{TB}$ phase (square electric pulses with duration of 10 μs, $E$ = 25 V/μm) [45], E-off- □ and E-on- ○.

For the relaxation times of the $N_{TB}$ distortion we obtain [7,10]:

$$\Gamma = 1/\tau_t = 2\pi f_t = \frac{2q^2 K_t \sin^2\theta}{\gamma} \qquad (6)$$

where: $K_t=(K_1+ K_2)/2$ is the effective elastic constant for the tilt in the coarse grained director $N$ away from the pseudo-layer normal. $\gamma$ is expected to be continuous at the N – $N_{TB}$ transition temperature and $\gamma$ is the rotational viscosity of the nematic phase. We can try to reproduce the temperature dependence of the relaxation rate by using experimentally available parameters for the CB7CB dimer using eq. (3). We used viscosity values obtained from the mode **m2**, see Fig. 1a, which is in good accordance with the bend viscosity in the range of N phase reported recently [44]. Wavenumber temperature dependence of the helix in the cell obtained from resonant carbon K-edge soft X-ray scattering by C. Zhu *et al* [21] and the cone angle data from birefringence measurements by C. Meyer *et al* [12]. Reliable data for elastic modulus are given for the CB7CB only in the range of N phase [44, 46]. Therefore $K_t$ was used in eq. (5) as an unknown parameter in fitting the predicted relaxation rate to the values of experimental results. The resulting elastic modulus is found to increase from 1 to 3 pN in the range of the $N_{TB}$ phase. It is assumed that the small measuring voltage (0.1V/μm) of the short period (~μs) does not change the pitch and associated with that the wavenumber of the heliconical structure. The results of the model prediction are quite striking. The resulting temperature dependence of the elastic modulus shows a clear softening behaviour at the transition temperature from the N to the $N_{TB}$ phase.

We can compare the above results with the relaxation rate calculated from the electro-optic experiment reported by C. Meyer et al [45], see Fig. 3. The electro-optic data initially shows a slight increase of the relaxation rate on decreasing temperature; then below 465 K the tendency reverses. However there overall the curve is almost horizontal and shows a much slower relaxation rate (~0.4 MHz) than what we found from relaxation data. The reason for this difference is intriguing. Indeed we note that the electro-optic experiment by C. Meyer et al [45] has been performed at reasonably high electric field amplitude of 25V/μm, nonetheless, the authors claim the field pulse was sufficiently short (10 μs) to consider the heliconical structure as conserved. We can test the strength of the influence of the bias field by applying a bias field (DC) in dielectric measurements. Fig. 4a shows the results for bias fields of 1 V/μm and 2 V/μm. With increasing the DC field the curves are becoming gradually less steep, which would indicate an unwinding of the heliconical structure, i.e. diminishing the wavenumber of the helix in the cell. This corresponds to pitch increases of about 20% and 36% for fields of 1 V/μm and 2 V/μm respectively. Thus, it is apparent r that the field applied in earlier electro-optic experiments can significantly deform the heliconical structure. We note that for $N_{TB}$ forming dimers in the presence of high magnetic fields a strong deformation and change of transition temperatures has been reported too [49]. For bias fields this unwinding effect in the presence of strong fields occurs despite the short switching period of the applied field. We observe that a periodic field is less efficient than a DC field; the corresponding pitch increases more than two times with respect to the off field helical structure. It has already been reported that an CB7CB sample in the $N_{TB}$ phase under an AC electric field, E=10 V/μm, parallel to the helix axis, a much higher birefringence $\Delta n$ [48], than without a field [12], Fig 4a. This suggests that an electric-field induced $N_{TB} - N_{SB}$ ($N_{SB}$ $_{SB =}$ splay bend) transition may take place. This effect has been as explained in the framework of the elastic-instability model [15,16] of the $N_{TB}$ and $N_{SB}$ phases by implying a strong biaxiality of the orientational order. However such a transition may also be an indication of the negative dielectric anisotropy of CB7CB in the range of $N_{TB}$ phase. In such a case the transition from the $N_{TB}$ to the $N_{SB}$ phase is expected at sufficiently strong electric fields [25,33]. The analysis of the data (see Fig 4a) points towards the field induced formation $N_{SB}$ phase, created below the N phase at sufficiently high bias fields. The lowest frequency relaxation process, **m4**, is observed in the frequency range from 1 Hz to $10^2$ Hz.

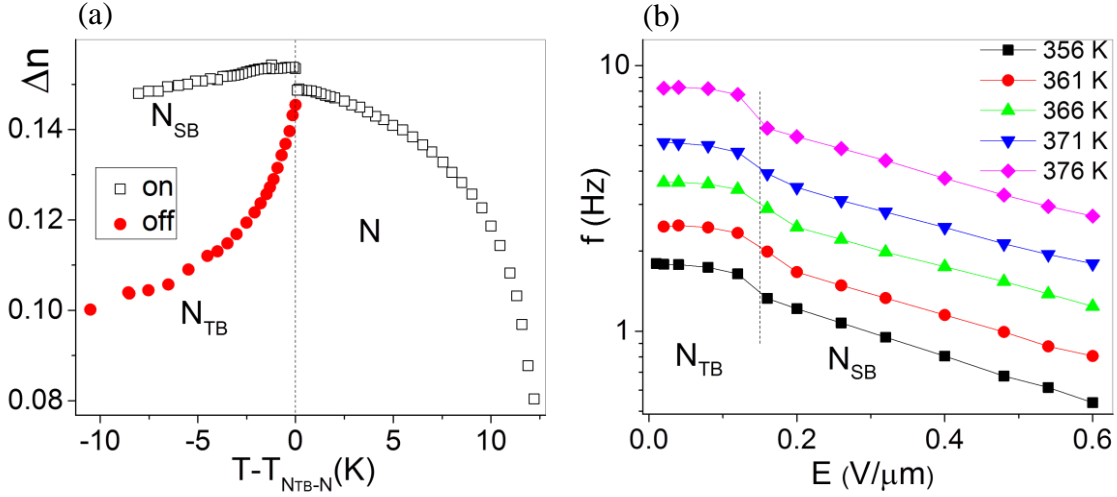

Fig.4.(a) Birefringence measurements [48]: field off ● and E=10 V/μm on -□. (b) Frequency of the low frequency mode f~$q^2$ plotted as a function of $E$, for temperatures: 351 K- ■, 356 K-●, 361 K-▲, 366 K-▼ and 376 K-♦.

This process can be identified as a Goldstone mode, related to long scale phase fluctuations of the heliconical director **n**, which does not change the coarse grained director *N* [10,25,32].
Such a long wavelength fluctuation mode should then be analogous to those found for the smectic-A phase (SmA), in particular the ''slow'', hydrodynamic layer compression. This mode should be described by an effective layer compression elastic constant $B_{eff}$ [10] (associated with $K_2$ and $K_3$) that plays the role of the elastic moduli for the compression of the pseudo-layer structure in the $N_{TB}$ phase. The low frequency dielectric results confirm the ''pseudo-layer'' structure of the $N_{TB}$ phase with an exceptionally low $B_{eff}$, values below $10^3$ Pa both for the CB7CB dimer and earlier observed for bent core LCs [25] and is somewhat lower than what has been reported for difluoroterphenyl based dimers forming the $N_{TB}$ phase [39]. The magnitude of $B_{eff}$ differs substantially from the typical value of $B_{eff} = 10^6$ Pa in a conventional SmA systems. Increasing the amplitude of the field up to 0.1 V/μm does not change the mode frequency, but with further increase the mode frequency drop down significantly and then decreases almost linearly. Assuming the mode frequency is proportional to $q^2$ [10], the drop of mode frequency at about 0.2 V/μm coincides with jump of wave number of the helical structure. Due to the model [33] this is again an indication of electric-field induced transition from $N_{TB}$ to $N_{SB}$ phase similar like for bent core sample [25].

**Conclusion**

In summary; we observe four modes in the dielectric spectra, the two higher frequency ones are assigned to the molecular modes and the other two are collective modes which are due to a distortion of the heliconical structure. The collective mode in the MHz region is assigned to local distortions of the conical angle, while the periodic helical structure remains unaltered. The temperature dependence of the relaxation frequency of the mode and resulting elastic modulus have anomalous, softening-like behaviour at the N-$N_{TB}$ transition. The dielectric method makes it possible to observe the modes without changing the period of the helical structure as opposed to the electro-optic methods. The lowest frequency collective mode, observed in the frequency range 1 Hz – $10^2$ Hz is particularly interesting and it can be identified as a Goldstone mode, related to long-scale fluctuation of the cone phase. This results in an alternating compression and expansion of the pseudo-layer structure. We interpret the data of the measurements in the presence of strong bias fields in the temperature range where typically the $N_{TB}$ phase is formed an indication of the formation of a field induced nematic splay bend phase ($N_{SB}$).

**Acknowledgements:** The author CW acknowledges funding by the EPSRC, grant EP/M015726/1.

The authors (K.M. and A.K.) thank the National Science Centre Poland for Grant No. 2011/03/B/ST3/03369.